\documentclass[conference]{IEEEtran}
\IEEEoverridecommandlockouts
\usepackage{cite}
\usepackage{amsmath,amssymb,amsfonts}
\usepackage{algorithmic}
\usepackage{graphicx}
\usepackage{textcomp}
\usepackage{xcolor}
\usepackage{subcaption}
\usepackage{multirow}
\usepackage{multicol}
\usepackage[top=0.75in, bottom=1.05in, left=0.625in, right=0.625in]{geometry}

\def\BibTeX{{\rm B\kern-.05em{\sc i\kern-.025em b}\kern-.08em
    T\kern-.1667em\lower.7ex\hbox{E}\kern-.125emX}}
    
\usepackage{fancyhdr}

\fancypagestyle{firstpage}{
  \fancyhf{} 
  \fancyhead[C]{\large Accepted to IEEE MILCOM 2025} 
}

\begin{document}

\title{An Adversarial-Driven Experimental Study on Deep Learning for RF Fingerprinting

\thanks{This research was supported in part by the National Science Foundation (NSF) under grants 2350255, 2218046, 2321271, and 2316720. Any opinions, findings, conclusions, or recommendations expressed in this paper are those of the author(s) and do not necessarily reflect the views of NSF.}
}


\author{
\IEEEauthorblockN{Xinyu Cao\IEEEauthorrefmark{1}$^1$~~Bimal Adhikari\IEEEauthorrefmark{1}$^2$~~Shangqing Zhao$^1$~~Jingxian Wu$^2$~~Yanjun Pan$^2$\\
\IEEEauthorrefmark{1}Co-First Authors\\
$^1$University of Oklahoma, Norman, OK\\
$^2$University of Arkansas, Fayetteville, AR\\
$^1${\em\{xinyu.cao-1,~shangqing\}@ou.edu}~~$^2${\em\{bimala,~wuj,~yanjunp\}@uark.edu}}
}
\maketitle
\thispagestyle{firstpage}

\begin{abstract}
Radio frequency (RF) fingerprinting, which extracts unique hardware imperfections of radio devices, has emerged as a promising physical-layer device identification mechanism in zero trust architectures and beyond 5G networks. In particular, deep learning (DL) methods have demonstrated state-of-the-art performance in this domain. However, existing approaches have primarily focused on enhancing system robustness against temporal and spatial variations in wireless environments, while the security vulnerabilities of these DL-based approaches have often been overlooked. In this work, we systematically investigate the security risks of DL-based RF fingerprinting systems through an adversarial-driven experimental analysis. We observe a consistent misclassification behavior for DL models under domain shifts, where a device is frequently misclassified as another specific one. Our analysis based on extensive real-world experiments demonstrates that this behavior can be exploited as an effective backdoor to enable external attackers to intrude into the system. Furthermore, we show that training DL models on raw received signals causes the models to entangle RF fingerprints with environmental and signal-pattern features, creating additional attack vectors that cannot be mitigated solely through post-processing security methods such as confidence thresholds.
\end{abstract}


\section{Introduction}

The proliferation of wireless devices and the expansion of beyond 5G networks have significantly increased the demand for reliable and secure device authentication methods. Radio frequency (RF) fingerprinting, which extracts the intrinsic hardware characteristics of radio devices, offers a promising physical-layer approach for device identification and authentication. RF fingerprinting is inherently difficult to forge or spoof, making it a valuable complement to traditional cryptographic-based security mechanisms. Due to its unique ability to distinguish devices based on their hardware signatures and its resistance to forgery, RF fingerprinting is envisioned as a key enabler for zero trust architectures and continuous device authentication and access control in 5G and beyond wireless networks \cite{jing2024authentication,siniarski2024robust}.

Recently, deep learning (DL)-based approaches have emerged as state-of-the-art techniques for RF fingerprint-based device identification, thanks to their superior performance and ease of deployment \cite{riyaz2018deep,sankhe2019oracle,reus2020trust,soltani2020more,restuccia2019deepradioid,gu2023attention,agadakos2020chameleons,li2022radionet}. Most existing research in this area focuses on improving system robustness against temporal and spatial variations in wireless environments. For example, multi-day training \cite{agadakos2020chameleons}, data augmentation \cite{soltani2020more}, artificial amplification of hardware features \cite{restuccia2019deepradioid,sankhe2019oracle}, novel neural network architectures  \cite{reus2020trust,gu2023attention}, and domain adaptation techniques such as transfer learning \cite{li2022radionet} have been introduced to improve robustness. Although these advances have strengthened the resilience of DL models for RF fingerprinting, the security aspects of these DL-based approaches have often been overlooked. Despite the fact that the RF fingerprint observed by a receiver is uniquely shaped by the pairwise characteristics of the transceiver pair, which makes it difficult to reproduce or replay, integrating DL models introduces a new attack surface that can potentially become the system’s weakest link.

\begin{figure}[t]
\centering
\begin{subfigure}[t]{0.48\linewidth}
  \includegraphics[width=\linewidth]{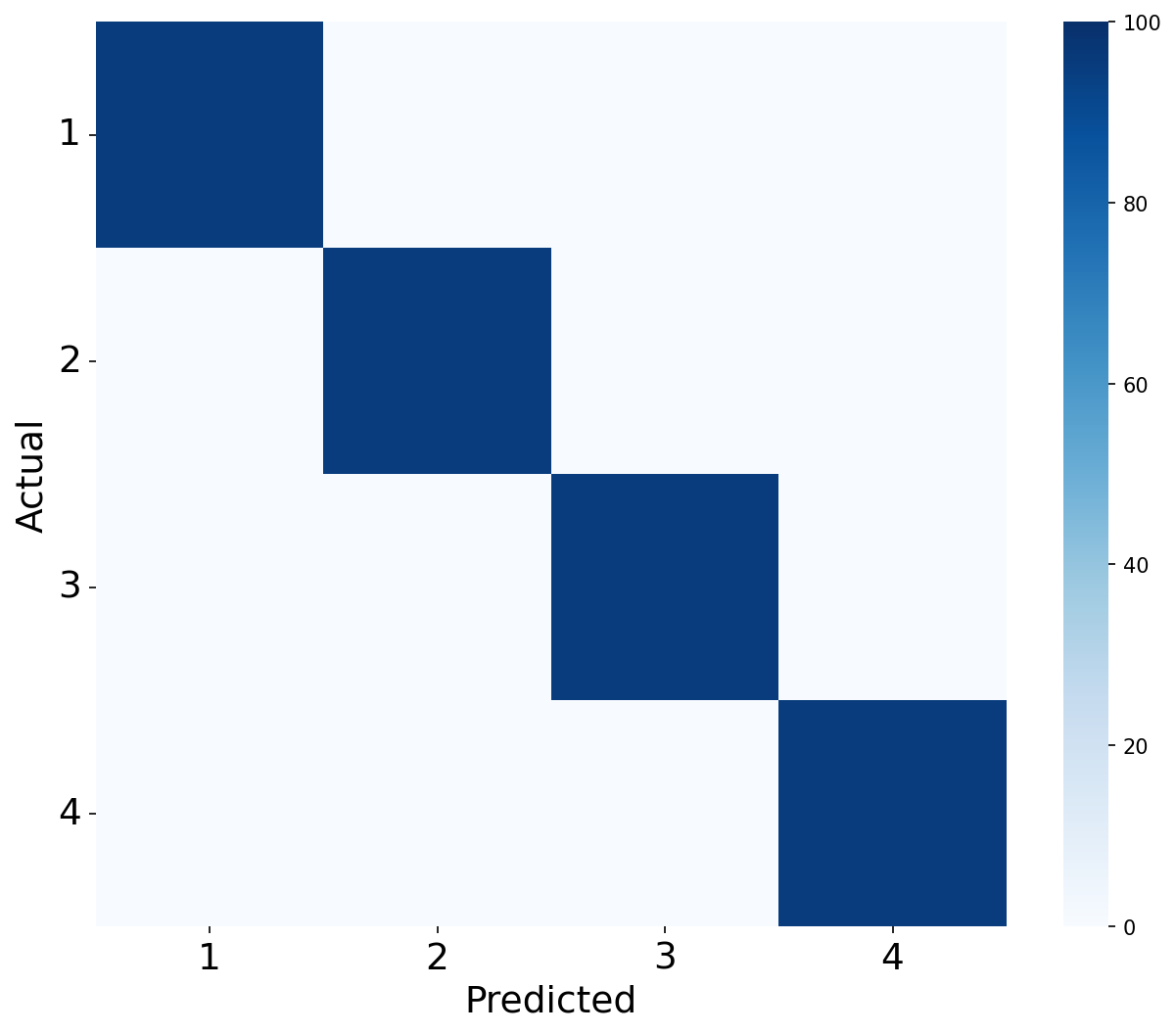}
  \caption{}
\end{subfigure}
\begin{subfigure}[t]{0.48\linewidth}
  \includegraphics[width=\linewidth]{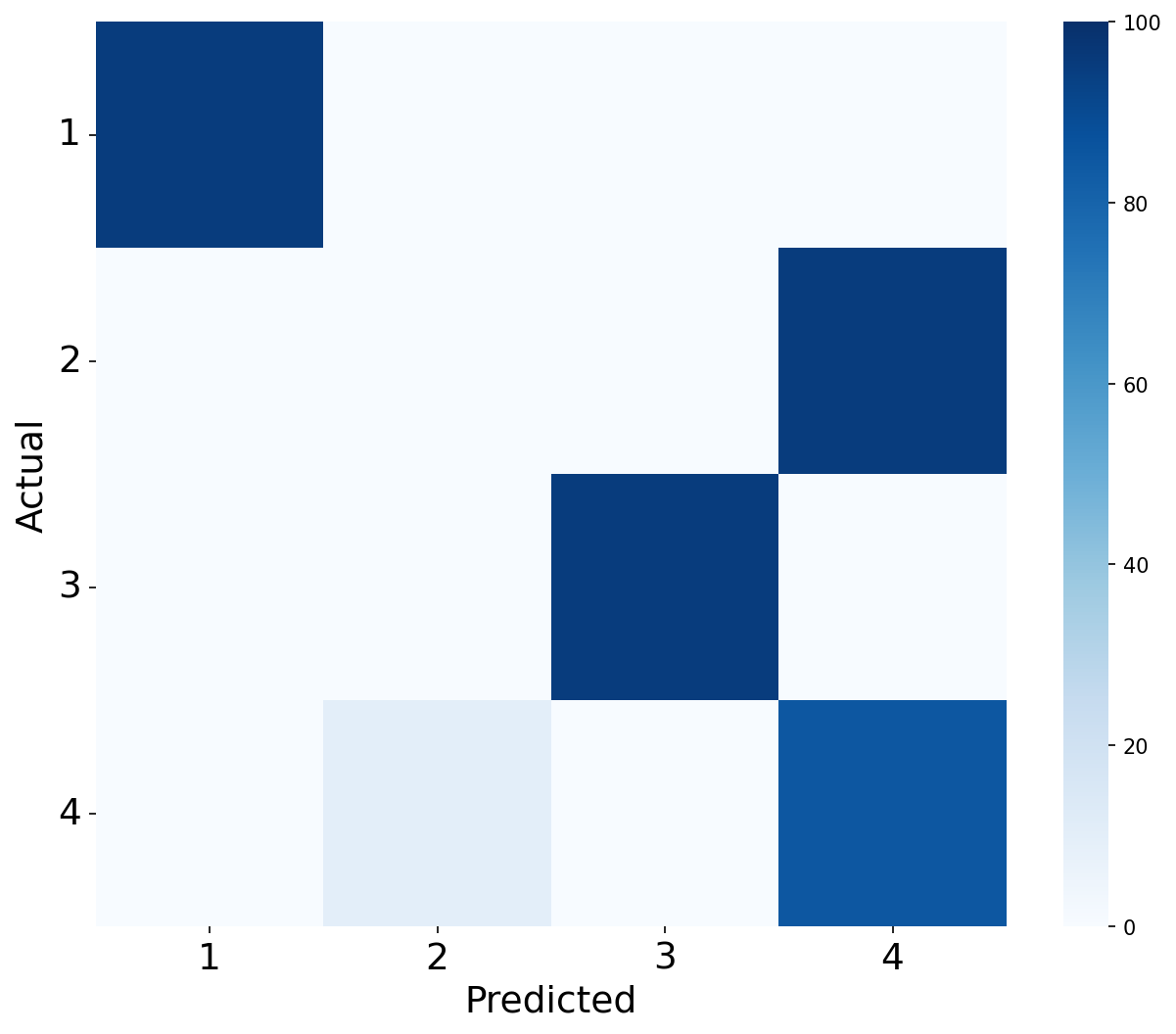}
  \caption{}
\end{subfigure}
\vspace{-5pt}
\caption{Illustration of typical performance of CNN-based device identification systems for four devices: (a) trained at time $t_1$ and location $l_1$; (b) tested at a different time ($t_2$) or location ($l_2$).}
\vspace{-15pt}
\label{fig:dl_motivation}
\end{figure}

In particular, we have observed a {\em consistent misclassification behavior} in such DL-based RF fingerprinting systems, where a given device is frequently identified as another specific one under domain shifts \cite{riyaz2018deep,sankhe2019oracle,reus2020trust,soltani2020more,restuccia2019deepradioid,gu2023attention,agadakos2020chameleons,li2022radionet}. Figure \ref{fig:dl_motivation} illustrates the typical performance of convolutional neural network (CNN)-based device identification systems using RF fingerprinting \cite{sankhe2019oracle,reus2020trust,soltani2020more,restuccia2019deepradioid,gu2023attention}. These systems commonly take raw received signals as input to a CNN trained to maximize classification accuracy. As shown in Fig. \ref{fig:dl_motivation}(a), the model performs effectively when evaluated on data collected at time ($t_1$) and location ($l_1$) as the training set. However, when tested on data gathered at a different time ($t_2$) or location ($l_2$), performance often degrades significantly. The confusion matrix in Fig. \ref{fig:dl_motivation}(b) reflects this degradation and the consistent misclassification behavior: while some devices may still be correctly identified with high confidence, others are consistently misclassified, often into specific incorrect classes. For instance, device 2 is consistently misclassified as device 4 with a high probability (often exceeding 90\%) in Fig. \ref{fig:dl_motivation}(b). This misclassification behavior occurs because CNNs tend to learn features that are highly specific to the training domain, making it difficult to generalize across variations in the data distribution caused by changes in time, location, or environmental conditions. From a security perspective, however, this misclassification pattern can be exploited as an unintended backdoor, potentially enabling impersonation attacks for external adversaries and thereby undermining the system’s security.

\begin{figure}[t]
\centering
\includegraphics[width=0.95\linewidth]{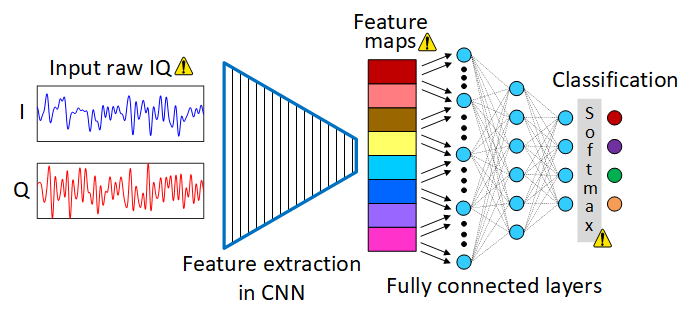}
\vspace{-5pt}
  \caption{The investigated CNN-based RF fingerprinting architecture.}
  \label{fig:overview}
  \vspace{-15pt}
\end{figure}

In this work, we fill a fundamental gap between the promises of RF fingerprinting and the practical realities of deploying DL models in dynamic, adversarial wireless environments through a comprehensive, adversarial-driven experimental analysis. We consider practical replay and naive impersonation attacks executed by external adversaries with no additional knowledge or technological advantage beyond standard equipment. The main contributions of this work are threefold and summarized as three key remarks, illustrated in Fig. \ref{fig:overview}.
\begin{itemize}
    \item To the best of our knowledge, we are the first to systematically evaluate DL-based RF fingerprinting systems from a security perspective. Unlike prior works that primarily focus on classification accuracy and domain robustness, we reveal a critical and underexplored vulnerability: the consistent misclassification behavior of CNN models under domain shifts can be exploited as an effective backdoor to enable external attackers to launch impersonation attacks and intrude into the system.
    \item Our extensive in-lab experimental results show that CNNs trained on raw received signals inadvertently entangle hardware-specific, hard-to-forge RF fingerprints with easy-to-reproduce environmental and signal-pattern features. This entanglement further renders the system highly vulnerable to impersonation attacks, even when the attacker lacks prior knowledge or channel control.
    \item Our evaluation of a commonly used security patch based on softmax confidence thresholding for CNNs demonstrates that it provides insufficient protection against impersonation attacks. Our findings highlight that the vulnerability stemming from the CNN's inability to disentangle hardware-specific features from spatiotemporal artifacts in wireless signals cannot be mitigated by post-processing techniques alone. Instead, carefully designed signal preprocessing methods are required to fully leverage the security potential of RF fingerprints.
\end{itemize}

The reminder of this paper is organized as follows: In Sec. II, we provide background and motivation for this work, followed by the system and threat models defined in Sec. III. Section IV describes the experimental setup, and Sec. V presents the security evaluation of the CNN-based RF fingerprinting system. We discuss potential future directions in Sec. VI and conclude the paper in Sec. VII.

\if 1
With the rapid proliferation of mobile and IoT devices, ensuring secure device authentication has become a fundamental requirement in wireless communication systems. Traditional cryptographic methods, while widely adopted, often rely on pre-shared keys or certificates, which are susceptible to theft, spoofing, or leakage. To address these limitations, RF fingerprinting has emerged as a promising physical-layer security solution that leverages the inherent hardware imperfections present in radio frequency (RF) transmissions. These imperfections, such as oscillator phase noise, power amplifier non-linearities, and IQ imbalance, are difficult to replicate precisely, providing a unique signature that can be used to identify and authenticate devices at the physical layer. 
\fi

\section{Background and Motivation}
\subsection{Input/Output Modeling with RF Fingerprints}
The transmitted data, denoted as $x[n]$, adheres to protocol standards and includes various components such as headers, preambles, and payloads. Due to unavoidable manufacturing imperfections in the transmitter's hardware, such as filters, oscillators, and clocks, the emitted signal subtly deviates from the ideal. These deviations result in unique, device-specific signal characteristics, known as RF fingerprints. Common RF fingerprints include carrier frequency offset (CFO), in-phase/quadrature (I/Q) imbalance, DC offset, phase noise, turn-on transients, and error vector magnitude (EVM) \cite{sankhe2019oracle}. These features are typically consistent over time for a given device and can be exploited for reliable identification. 

Considering the effects of CFO, phase offset, and IQ imbalance at both the transmitter and receiver, as well as the wireless channel, and assuming that the $L$-length channel is quasi-static over the duration of an OFDM symbol (i.e., allowing the channel to be modeled as time-invariant within the symbol duration) we can represent the time-domain baseband signal of an OFDM symbol at the receiver as \cite{tandur2006joint}
\begin{align}
y[n] & = \eta e^{-j(n\Delta\omega_{RT} T_s + \psi_{RT})} r[n] \notag \\
& ~~+ \kappa e^{j(n\Delta\omega_{RT} T_s + \psi_{RT})} r^*[n] + w[n]
\label{eq:y_n}
\end{align}
where $n = 0, 1, \dots, N-1$, $T_s$ is the OFDM sampling period of the OFDM system, and $w[n]$ is additive white Gaussian noise. $r[n] = \sum\limits_{l=0}^{L-1} h[l] x[n-l]$, the composite time-varying CSI $h[l] = p_T(nTs) \otimes g(nTs, lT_s) \otimes p_R(nTs)$ includes the transmit filter $p_T(t)$, the receive filter $p_R(t)$, and the time-varying impulse response of the physical fading channel $g(t, \tau)$. The composite CFO is $\Delta \omega_{RT} = \Delta \omega_R - \Delta \omega_T$, and the composite phase offset is $\psi_{RT} = \psi_R - \psi_T$. The parameters $\eta = (\alpha_R \alpha_T + \beta_R \beta_T^*)$, $\kappa = (\alpha_R\beta_T+\beta_R\alpha_T^*)$, where $\alpha_{(\cdot)}$ and $\beta_{(\cdot)}$ represent the effects of IQ imbalance of either the transmitter or the receiver. W.l.o.g, for the transmitter $\alpha_T = \cos(\Delta\phi_T) + j\epsilon_T\sin(\Delta\phi_T)$ and $\beta_T = \epsilon_T\cos(\Delta\phi_T) - j\sin(\Delta\phi_T)$, where $\epsilon_T$ and $\Delta\phi_T$ represent the amplitude and phase differences between the transmitter's IQ branches. 

Equation \eqref{eq:y_n} shows that the RF fingerprint observed by a receiver such as  an access point (AP) is uniquely shaped by the pairwise characteristics of the transceiver pair. As a result, it is generally considered difficult to reproduce or replay RF fingerprints. This is because any replaying device introduces its own hardware impairments, which distort the composite RF fingerprint perceived by the AP. Hence, the replay attack is widely believed to have limited effectiveness.



\subsection{Vulnerability Analysis}
State-of-the-art DL-based RF fingerprinting and device identification approaches have primarily focused on enhancing system robustness against temporal and spatial variations in wireless environments \cite{sankhe2019oracle,reus2020trust,soltani2020more,restuccia2019deepradioid,gu2023attention,agadakos2020chameleons,li2022radionet}. However, these works often neglect thorough security evaluations of the systems themselves. Although the inherent uniqueness of RF fingerprints makes forging another device's signal challenging, the DL model can become the system’s weakest link.
Specifically, we observed a {\em consistent misclassification behavior} in DL models, where a device is frequently misclassified as another specific device under domain shifts, as illustrated in Fig. \ref{fig:dl_motivation}(b). This behavior can unintentionally introduce exploitable vulnerabilities. Despite the physical uniqueness of RF fingerprints making forgery difficult, this recurrent misclassification, shown in several prior works \cite{sankhe2019oracle,reus2020trust,soltani2020more,restuccia2019deepradioid,gu2023attention}, creates a new attack surface that adversaries may exploit for impersonation attacks.
This vulnerability exposes a critical gap in current DL-based RF fingerprinting research, i.e., the security implications of the learned feature representations remain insufficiently explored. To address this gap, in this work, we conduct a comprehensive, adversarial-driven experimental analysis by considering two straightforward and practical yet later proven effective attacks.

\if 1
Recently, deep learning (DL)-based approaches have emerged as the state-of-the-art in RF fingerprint-based device identification systems, due to their ease of deployment and superior performance \cite{riyaz2018deep,sankhe2019oracle,reus2020trust,li2022radionet,gu2023attention}. Most existing research in this area focuses on improving the system’s robustness against temporal and spatial variations in wireless environments \cite{reus2020trust,li2022radionet,gu2023attention}. However, these studies often neglect a comprehensive evaluation of the system’s security. While the inherent uniqueness of RF fingerprints makes it challenging to forge another device's signal, the DL model itself may become the system’s weakest link.

Figure \ref{fig:dl_motivation} illustrates the typical performance of CNN-based device identification systems using RF fingerprinting \cite{sankhe2019oracle,reus2020trust}. These systems commonly use raw IQ samples, as modeled in Eq. \eqref{eq:y_n}, as input to a CNN trained to maximize classification accuracy. As shown in Fig. \ref{fig:dl_motivation}(a), the model performs effectively when evaluated on data collected at the time ($t_1$) and location ($l_1$) as the training set. However, when tested on data gathered at a different time ($t_2$) or location ($l_2$), performance often degrades significantly. The confusion matrix in Fig. \ref{fig:dl_motivation}(b) reflects this degradation: while some devices may still be correctly identified with high confidence, others are consistently misclassified, often into specific incorrect classes. This degradation occurs because CNNs tend to learn features that are highly specific to the training domain, making it difficult to generalize across variations in the data distribution caused by changes in time, location, or environmental conditions. As a result, recent research in DL-based RF fingerprinting has focused on enhancing model robustness and developing domain adaptation techniques to maintain performance in conditions that differ from those seen during training.

On the other hand, RF fingerprint-based device identification systems are designed to authenticate each device using its unique hardware-induced characteristics. However, while much of the prior work has focused on improving classification accuracy and domain robustness, the security aspects of these DL-based approaches have often been overlooked. Notably, the {\em consistent misclassification behavior} of DL models, where a given device is frequently identified as where a device is frequently misclassified as another specific one under domain shifts, can unintentionally create exploitable vulnerabilities. Although the physical uniqueness of RF fingerprints makes them inherently difficult to forge, such predictable misclassification behavior can, from the perspective of a DL model, function as an unintended backdoor. This opens the door to potential impersonation attacks, thus undermining the system’s security.
\fi

\section{System and Threat Models}
This section defines the system and threat models, including the two practical attacks that we use to evaluate the security of DL-based RF fingerprinting and device identification systems.
\subsection{RF Fingerprint-based Device Identification Systems}

RF fingerprint-based device identification and management systems typically involve a receiver, usually a radio AP, that aims to identify and authenticate any transmitter attempting to establish a communication link. For instance, in  5G and beyond wireless networks, the base station (BS), which acts as a fixed receiver and serves as the AP to the network, must authenticate any user equipment (UE) attempting to connect. This process involves two main phases: fingerprinting and identification.

During the fingerprinting phase, the AP requires each transmitter to send signals in accordance with the communication protocol for a certain period during its first access attempt to the network. The AP can either process the received signal to explicitly extract RF fingerprints for input into a classification model, or directly input the received IQ samples into a model for classification. In this work, we adopt the second approach as in state-of-the-art systems, the AP directly feeds the raw IQ signals into a DL model such as a CNN for fingerprinting \cite{sankhe2019oracle,reus2020trust,li2022radionet,gu2023attention}. An illustration of the investigated CNN-based RF fingerprinting architecture is shown in Fig. \ref{fig:overview}.

During the identification phase, when a transmitter seeks to connect to the AP, the AP collects its transmitted radio signals and performs identification with the pretrained DL model for RF fingerprinting. Based on the classification outcome, the AP either authenticates the device and permits connection establishment, or denies access. 

\subsection{Threat Model}
We consider an external attacker attempting to compromise an RF fingerprint-based device identification system. The attacker’s goal is to impersonate an authorized device in order to gain unauthorized access to the network. We assume a practical attacker with no additional knowledge or technological advantage beyond standard equipment. The attacker is located  away from the AP and legitimate transmitters.

The attacker’s capabilities are limited to passively collecting signals emitted by the legitimate transmitters over the air during both the fingerprinting and identification phases. The attacker then either replays these captured signals or transmits synthesized signals from its own transmitter to bypass the DL-based device identification system. The attacker cannot control the communication channel between the AP and any legitimate transmitter, nor inject false data to disrupt the AP’s model training. We consider the following two attack scenarios:

$\bullet$ {\em Replay attack:} The attacker records signals emitted by legitimate transmitters during the fingerprinting phase and replays them during the identification phase to fool the DL-based authentication system.

$\bullet$ {\em Naive impersonation attack:} During the identification phase, the attacker simply transmits signals that mimic the format used by legitimate devices during CNN model training (e.g., preambles or random payloads) using its own transmitter.

\if 1
Assuming a fixed transmitted signal and AWGN with constant variance, RF fingerprint-based device authentication primarily relies on device-specific hardware impairments and channel-dependent propagation effects as shown in Eq. \eqref{eq:y_n}. There are two main reasons why machine learning models may struggle to effectively extract device-specific features in this context.

First, due to the multiplicative and highly non-linear nature of the received signal, accurately isolating hardware-induced features becomes a complex non-convex optimization problem. Standard neural networks, which are typically designed to learn approximately linear or piecewise linear mappings, are not well suited to extract these features directly, especially in the absence of carefully designed, domain-specific signal preprocessing techniques. Consequently, the model can fail to capture the correct latent representations associated with hardware impairments.

Second, in practical wireless environments, the magnitude and variability of channel-dependent effects often overshadow those caused by hardware impairments. For instance, multipath fading can introduce power fluctuations in the range of –5 to +5 dB, while impairments like IQ imbalance and CFO are typically much smaller (e.g., $\epsilon \in [0.01, 0.05]$ and $\Delta f \in [50, 1000]$ Hz) \cite{xxx}. When raw IQ samples are used as input without explicitly separating these effects, preprocessing techniques such as normalization cannot correct for the discrepancy in their relative strengths. As a result, the dominant channel effects can mask the subtler device-specific signatures. Given that machine learning models are sensitive to the scale and variance of input features, this imbalance can significantly degrade generalization performance of the model, especially under dynamic or previously unseen channel conditions.

\subsection{Motivation of Attack Designs}

Assuming a fixed transmitted signal and AWGN with constant variance, RF fingerprint-based device authentication relies primarily on the multiplicative effects imposed on the received signal. We model the effective multiplicative coefficient applied to the transmitted signal as a composition of channel-dependent propagation effects and device-specific hardware impairments, expressed as
\begin{equation}
\begin{split}
  &h(t) = \underbrace{\left( \sum_{k=1}^{L} \alpha_k(t) e^{-j 2\pi f_{D,k} t} e^{-j \theta_k} \right)}_{\text{Channel-dependent effects: } h_c(t)} \\
  &\times \underbrace{\left( e^{j(2\pi \Delta f t + \phi_0)} \left[ (1 + \epsilon_I) + j(1 + \epsilon_Q) \cdot \rho(t) \right] \right)}_{\text{Device-specific impairments: } h_{\text{dev}}(t)}.
\end{split}
\end{equation}
Here, the first term $h_c(t)$ captures the wireless channel effects, including multipath fading represented by $\alpha_k(t)$, Doppler shifts $f_{D,k}$, and phase rotations $\theta_k$, all of which are inherently time-varying and location-dependent. The second term $h_{\text{dev}}(t)$ captures device-specific impairments such as CFO $\Delta f$, initial phase offset $\phi_0$, IQ gain imbalance $(\epsilon_I, \epsilon_Q)$, and quadrature distortion modeled by $\rho(t)$. 

There are two primary reasons why machine learning models may fail to effectively capture device-specific features for RF fingerprint-based authentication. First, due to the complex multiplicative nature of the received signal, accurately isolating the device-specific components from this composite transformation generally requires solving a non-convex optimization problem, often involving NP-hard formulations due to the nonlinear and non-separable nature of the underlying signal model (e.g., the imbalance factor $\epsilon$ and phase offset $\theta$ are highly coupled through nonlinear relationships). Standard neural networks, which are typically optimized for learning approximately linear or piecewise linear mappings, are  not suited to directly recover such features simultaneously without explicit domain-specific signal preprocessing. As a result, the model may fail to learn the correct latent representations corresponding to hardware-induced impairments.

Second, in practical wireless environments, the amplitude and variability of the channel-dependent term $h_c(t)$ often dominate those of the device-specific term $h_{\text{dev}}(t)$ by one or two orders of magnitude. For example, multipath fading can cause power fluctuations in the range of $-5$ to $+5$ dB, whereas hardware impairments such as IQ imbalance and CFO generally fall within a few percent (e.g., $\epsilon_I, \epsilon_Q \in [0.01, 0.05]$ and $\Delta f \in [50, 1000]$ Hz). When raw IQ samples are directly fed into a neural network without explicitly separating or normalizing these contributions, the dominant channel effects tend to overshadow the relatively weak device-specific features. Since machine learning models are sensitive to the scale and variance of their input features, this imbalance leads to poor generalization and degraded authentication performance, particularly in the presence of dynamic or previously unseen channel conditions.

In a typical RF fingerprint-based authentication scenario, consider two mobile users, Alice and Bob, who both attempt to establish communication with a base station. Each device transmits a known pilot signal $s(t)$, but due to unavoidable manufacturing variances and hardware impairments, the transmitted signals differ subtly. The actual transmitted signal from each device, denoted as $x_i(t)$ where $i$ represents either Alice or Bob, can be mathematically modeled as:
\begin{equation}\label{Eq:rx_signals}
x_i(t) = h_i(t) \cdot s(t) + n_i(t),
\end{equation}
where $h_i(t)$ captures the device-specific RF fingerprint characteristics, while $n_i(t)$ accounts for additive transmitter noise. These RF fingerprints embedded in $h_i(t)$ include features such as carrier frequency offset (CFO), amplitude and phase mismatches, and spectral regrowth, which can be exploited by the receiver for device identification.

Upon reception, the signal $x_i(t)$ is down-converted to baseband, resulting in complex in-phase and quadrature (IQ) samples, denoted as $x_i[n]$, where $n$ is the discrete-time index. These IQ samples inherently carry the device-specific RF impairments introduced by $h_i(t)$, in addition to the channel and noise effects. Then, rather than explicitly extracting handcrafted features from the received signals, the raw IQ samples themselves are directly used as inputs to a machine learning (ML) model to automatically learn discriminative patterns from data. Mathematically, the raw IQ sample sequence is represented as:
\begin{equation}\label{Eq:IQ_seq}
\mathbf{x}_i = [x_i[1], x_i[2], \ldots, x_i[N]],
\end{equation}
where $N$ is the length of the collected sample window. A classifier $\mathcal{C}$ is trained using a historical dataset of labeled IQ sequences, denoted as $\mathcal{D} = \{ (\mathbf{x}_i, y_i) \}$, where $y_i \in \{\text{Alice}, \text{Bob}\}$. The objective of the classifier is to learn the mapping:
\begin{equation}
\hat{y} = \mathcal{C}(\mathbf{x}_i),
\end{equation}
where $\hat{y}$ is the predicted device identity. In literature, typical models $\mathcal{C}$ used for this purpose include convolutional neural networks (CNNs), recurrent neural networks (RNNs), and hybrid deep learning architectures, which are capable of capturing both spatial and temporal dependencies present in the IQ sequences.

\subsection{Vulnerability Analysis}

Although RF fingerprinting methods that directly leverage the raw IQ samples $\mathbf{x}_i$ for device classification have demonstrated promising accuracy in controlled environments, they exhibit critical vulnerabilities when applied in realistic, time-varying wireless channels. As previously described in \eqref{Eq:rx_signals}, $h_i(t)$ encapsulates the device-specific RF fingerprint, however, in practical scenarios, $h_i(t)$ is not isolated; instead, it is compounded with the wireless environment context $c(t)$, which is implicitly embedded in the received signal 
\begin{equation}
    x_i(t) = c(t) \cdot h_i(t) \cdot s(t) + n_i(t) \label{eq:rawIQ}
\end{equation}
Upon downconversion to baseband, the complex IQ samples $x_i[n]$ reflect both the device-specific hardware impairments and the dynamic wireless channel effects introduced by $c(t)$. Consequently, when these raw IQ sequences \eqref{Eq:IQ_seq} are directly used as the input to the machine learning classifier $\mathcal{C}(\cdot)$, the model learns a composite representation that implicitly couples the static device fingerprint $h_i(t)$ and the highly time-variant channel characteristics $c(t)$.

This observation creates a mismatch with the intrinsic design philosophy of machine learning models, which are fundamentally optimized for recognizing static or quasi-static patterns. As the wireless channel $c(t)$ changes over time due to environmental dynamics (e.g., mobility, obstacles, multipath), the composite signal characteristics within $\mathbf{x}_i$ also shift, despite the underlying device-specific fingerprint $h_i(t)$ remaining unchanged. Therefore, the learned decision boundaries of the classifier $\mathcal{C}(\cdot)$, which are anchored on the training data collected at a specific time and channel state, may no longer generalize well when applied to testing data collected at a later time under different channel conditions.


\subsection{Motivation of Attack Designs}
As illustrated in Eq. \eqref{eq:rawIQ}, the raw IQ signals used as training data in deep learning based RF fingerprinting approaches inherently embed not only the device-specific RF fingerprint $h_i(t)$ but also the environmental context $c(t)$ including objects, walls, and other surfaces that cause multipath effects and fading and are unique to a given wireless setting. While deep learning based RF fingerprinting models are designed to learn discriminative device-specific patterns, this entanglement with the environmental context introduces a critical and underexplored vulnerability, that is, the learned model may inadvertently overfit to the public and easily cloned environmental characteristics rather than invariant device-specific features. Existing studies have shown that models trained on data collected under one environmental condition (e.g., Day 1) suffer notable degradation in classification performance when evaluated under slightly altered conditions (e.g., Day 2), even when the same devices are used \cite{}. This behavior indicates a concerning over-reliance on the environmental component $c(t)$, which is public and easily cloned in real-world deployments.

Motivated by this insight, we explore a new class of adversarial threats rooted in the environment-induced bias of RF fingerprinting models. Specifically, we consider an attacker situated in an environment similar to that of the BS during the training phase, capable of passively recording legitimate transmissions. Such an adversary can later replay these or carefully crafted signals in a manner that preserves environmental features similar to those present during training, thereby increasing the likelihood of being misclassified as a legitimate transmitter. Unlike traditional spoofing attacks that attempt to replicate device-specific and unclonable RF features directly and often require precise hardware manipulation, our proposed attack leverages a more accessible vector of environmental similarity bias. In addition, this attack mechanism is both feasible and stealthy. It does not require detailed knowledge of the deep learning architecture or parameters, nor does it demand physical access to the legitimate device. Instead, it exploits the very foundation of modern deep learning models in RF systems—namely, their reliance on high-dimensional correlations in the training data, which, in this case, include environmental artifacts unintentionally treated as discriminative features.
\fi

\section{In-Lab Experimental Setup}
We conduct an adversarial-driven experimental study on a CNN-based device identification system using IQ samples collected from an in-lab testbed. The testbed consists of 6 USRP X300 SDRs, each equipped with a UBX-160 daughterboard operating at a center frequency of 5.78 GHz and a sampling rate of 192 KHz. In each experiment, four of these USRPs act as legitimate transmitters (TXs), one as the attacker (Eve), and one as the AP. All devices are connected to the same host computer equipped with an AMD Ryzen 9 9950X CPU, 96 GB RAM, 6 TB storage, and an NVIDIA RTX 4090 GPU.

All transmitters emit IEEE 802.11a-compliant OFDM frames using QPSK modulation, generated via GNU Radio. Each transmitted frame has a fixed length and contains three OFDM symbols, including a header with CRC, a cyclic prefix, and payload data. Two types of frames are used: those with fixed payloads (referred to as repeated signals) and those with random payloads (referred to as random signals). The AP is a fixed-endpoint USRP that collects raw IQ samples from all transmitters. 
Eve is another fixed-endpoint USRP positioned approximately 2m away from the AP. During the fingerprinting phase, when the AP collects signals from each transmitter, Eve simultaneously records the raw IQ samples she receives.

All experiments are conducted in an indoor lab environment under non-line-of-sight (NLoS) conditions only, i.e., there is no direct path between any transmitter–receiver pair. The distances between transmitter–receiver pairs vary and can extend up to 18m. This NLoS setting introduces significant multipath fading and spatiotemporal channel fluctuations, providing sufficient data diversity for CNN training.

In each experiment, we collect three sets of 8 million IQ samples per transmitter at both the AP and Eve. This entire collection process is repeated two hours later. The initial dataset is used for training, while the second dataset is reserved for testing. The datasets used for system performance evaluation are categorized as the training set (TrS), the testing set (TeS), and the attacking set (AS), as detailed below.

$\bullet$ TrS 1: Collected on Day 1 at time $t_1$, with TXs sending random signals at location $l_1$.

$\bullet$ TeS 1: Collected on Day 1 at time $t_2$, with TXs sending random signals at location $l_2$.

$\bullet$ TrS 2: Collected on Day 2 at time $t_3$, with TXs sending repeated signals at location $l_3$.

$\bullet$ TeS 2: Collected on Day 2 at time $t_4$, with TXs sending repeated signals at location $l_4$.

$\bullet$ AS 1: Collected by Eve alongside TrS 1.

$\bullet$ AS 2: Collected by Eve alongside TrS 2.

The training and testing sets are collected by the AP, while the attacking sets are collected by Eve. Each training dataset is split into 70\% for training and 30\% for validation, while each testing dataset is used entirely for testing.

\if 1
\textcolor{blue}{Input from Shangqing/Xinyu} In this section, we conducted mutiple times experiments  using 6 USRP X310 SDRs (each with a UBX-160 daughterboard operating at a center frequency of 5.78 GHz with 192 kHz sampling rate): 4 as legitamate transmitters (TXs), 1 as eavesdropper (EVE) and 1 as  Base Station (BS). All devices are connected to the same computer host equipped with AMD Ryzen 9 9950X, 96 GB RAM, 6TB storage and NVIDIA RTX 4090.

A fixed endpoint USRP serves as the BS  that authenticates and connects with available transmitters. It collects the raw IQ samples from the transmitters and runs the inference model to verify the TX ID. Each transmitter sends a uniformly random payload using the same OFDM-QPSK modulation. The other fixed endpoint USRP servers as the EVE. We set the EVE at the center frequency of TXs, thus it has the capability to collect the RAW IQ samples emitted by TXs then replay the collected samples to BS to launch the attack.

Raw IQ samples are collected in an indoor environment over purely non-line-of-sight (NLOS) links, with transmitter-receiver separations extending up to 18 m. The absence of a direct path in each link created noticeable multipath fading and time-varying channel fluctuations. At each collection point, we collect 3 sets of 8 million IQ samples per TX at the BS, then repeat the entire collection process 2 hours later. Each frame of collected IQ samples was transposed into sequences of 192 samples. The initial dataset was divided into 70\% for training, 30\% for validation. The second dataset is used entirely for model testing.
\fi

\section{Performance Evaluation}
In this section, we evaluate the security of the CNN-based RF fingerprinting system for device identification against replay and naive impersonation attacks. 


\subsection{Classifier Architecture}
In this work, we adopt the CNN architecture presented in \cite{reus2020trust}, which is a typical design consisting of eight layers including four convolutional layers followed by three fully connected (dense) layers. All input data are normalized using per-device standardization, where for each deice, we flatten its I/Q data and fit using only that device's data before being fed into the CNN model.
For the input, a sliding window of frame size 4 is implemented that makes the input as 2$\times$256 where the I and Q components of the signal are stored in two rows of matrix. Each of the convolutional layers uses 40 filters, where the first two have kernel sizes of 1$\times$7 and 1$\times$5, respectively, each followed by a max-pooling layer that reduces the dimensionality to 2$\times$128$\times$40 and 2$\times$64$\times$40, respectively. These are followed by a convolution layer of size 2$\times$7 and max-pooling layer that further reduces the dimension to 2$\times$32$\times$40. Finally, a convolution layer of size 2$\times$5 is implemented followed by a flattening layer, and the final feature maps are passed through three dense layers with 1024, 256 and 4 neurons, respectively. Dropout is applied after the flattening and first dense to help prevent overfitting. The final classification layer uses softmax activation to output probabilities over 4 classes. We use Adam optimizer with a learning rate of 1$\times10^{-3}$ to train the model.


\subsection{Experiment Results}
As Eq. \eqref{eq:y_n} shows, the received signal is a complex and nonlinear combination of multiple factors, including RF fingerprints, the time- and space-varying wireless channel, the transmitted signal, and AWGN. Although the CNN is intended to learn hardware-specific RF fingerprints only, when trained directly on raw IQ data, standard neural networks, which are typically designed to learn approximately linear or piecewise linear mappings, are not well suited to extract these features directly, especially in the absence of carefully designed, domain-specific signal preprocessing techniques. Hence, the model may fail to capture the correct latent representations associated with hardware impairments. To facilitate the analysis of system performance and security, we consider two types of transmitted frames, as described in Section IV. In the following, we evaluate the system using IQ samples collected under each of these two frame types, respectively.

\subsubsection{Consistent misclassification behavior during domain shifts} 
\begin{figure}[t]
\centering
\begin{subfigure}[t]{0.48\linewidth}
  \includegraphics[width=\linewidth]{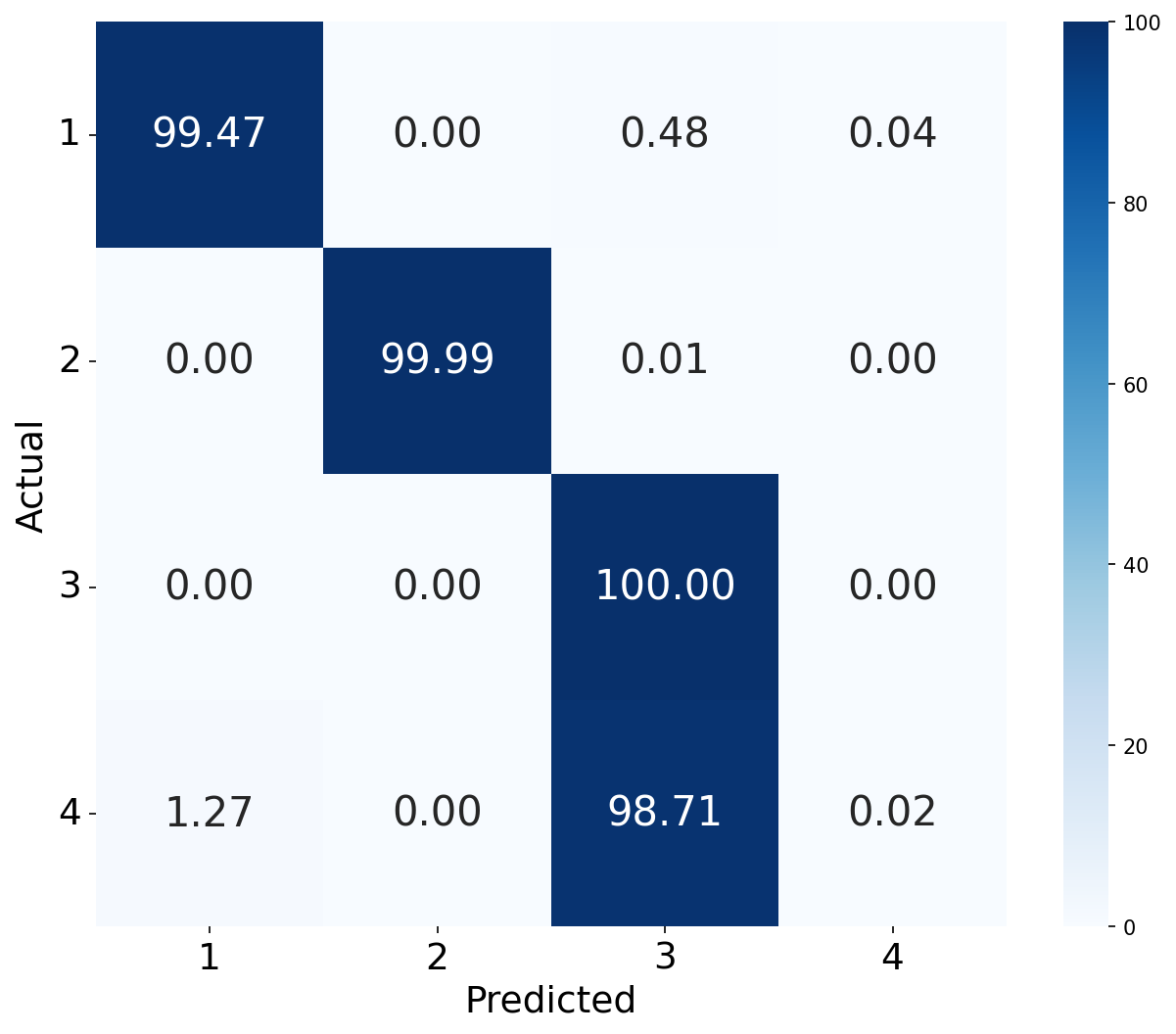}
  \caption{}
\end{subfigure}
\begin{subfigure}[t]{0.48\linewidth}
  \includegraphics[width=\linewidth]{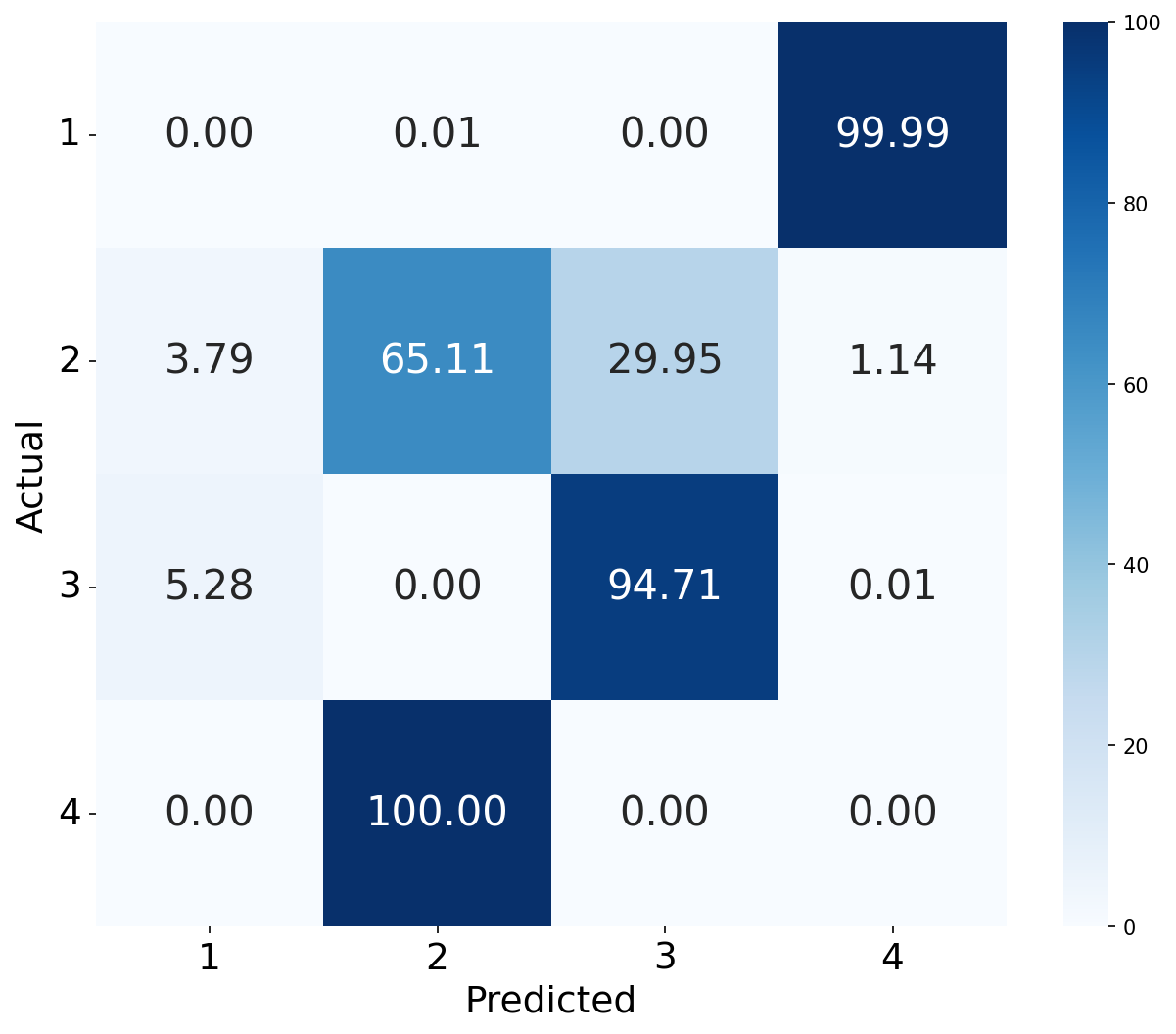}
  \caption{}
\end{subfigure}
\vspace{-5pt}
\caption{Confusion matrix showing the classification accuracy of the CNN model trained on TrS 1: (a) tested on TeS 1; (b) tested on AS 1.}
\vspace{-10pt}
\label{fig:ex_differ}
\end{figure}

We begin with the CNN trained using TrS 1 and test it on TeS 1 and AS 1, respectively, as existing works typically assume transmitters send random payloads \cite{riyaz2018deep,sankhe2019oracle,reus2020trust,soltani2020more,restuccia2019deepradioid,gu2023attention,agadakos2020chameleons,li2022radionet}. Figure \ref{fig:ex_differ}(a) shows that the device authentication accuracy of the trained model experiences a significant drop when tested on TeS 1 due to the domain shift, compared to the overall training accuracy of 99.95\%. In other words, when there is a domain shift between the training and testing sets, we observe a misclassification behavior of the CNN similar to that reported in the literature. For example, device 4 is frequently identified as device 3 with a probability as high as 98.71\%.

To verify whether this misclassification behavior of the model during domain shifts is consistent and allows Eve to easily penetrate the system by impersonating a legitimate transmitter, we show the success rate of the replay attack launched by Eve. In this untargeted attack, Eve replays AS 1 to impersonate any legitimate device. As illustrated in Fig. \ref{fig:ex_differ}(b), we observe that Eve can successfully impersonate Devices 2, 3, and 4 with relatively high probabilities ($\geq$ 95\%) and an overall untargeted impersonation attack success rate of 95.96\%.

Prior works attribute the consistent misclassification behavior of the CNN to the similarity in wireless channels experienced by certain transmitter pairs. However, since the correlation coefficients of the channel state information (CSI) for each communication pair across TrS 1, TeS 1, and AS 1 are low (significantly below 0.5, as shown in Fig. \ref{fig:corr}), yet the misclassification behavior remains consistent, we conclude that this behavior is an inherent limitation of the CNN, caused by its lack of generalization during domain shifts.

{\bf Remark 1.} {\em The consistent misclassification behavior of the CNN-based device identification system is caused by the lack of generalization of the model during domain shifts, leaving the system vulnerable to untargeted impersonation attacks.}

\subsubsection{Entangled feature maps} 
\begin{figure}[t]
\centering
\begin{subfigure}[t]{0.48\linewidth}
  \includegraphics[width=\linewidth]{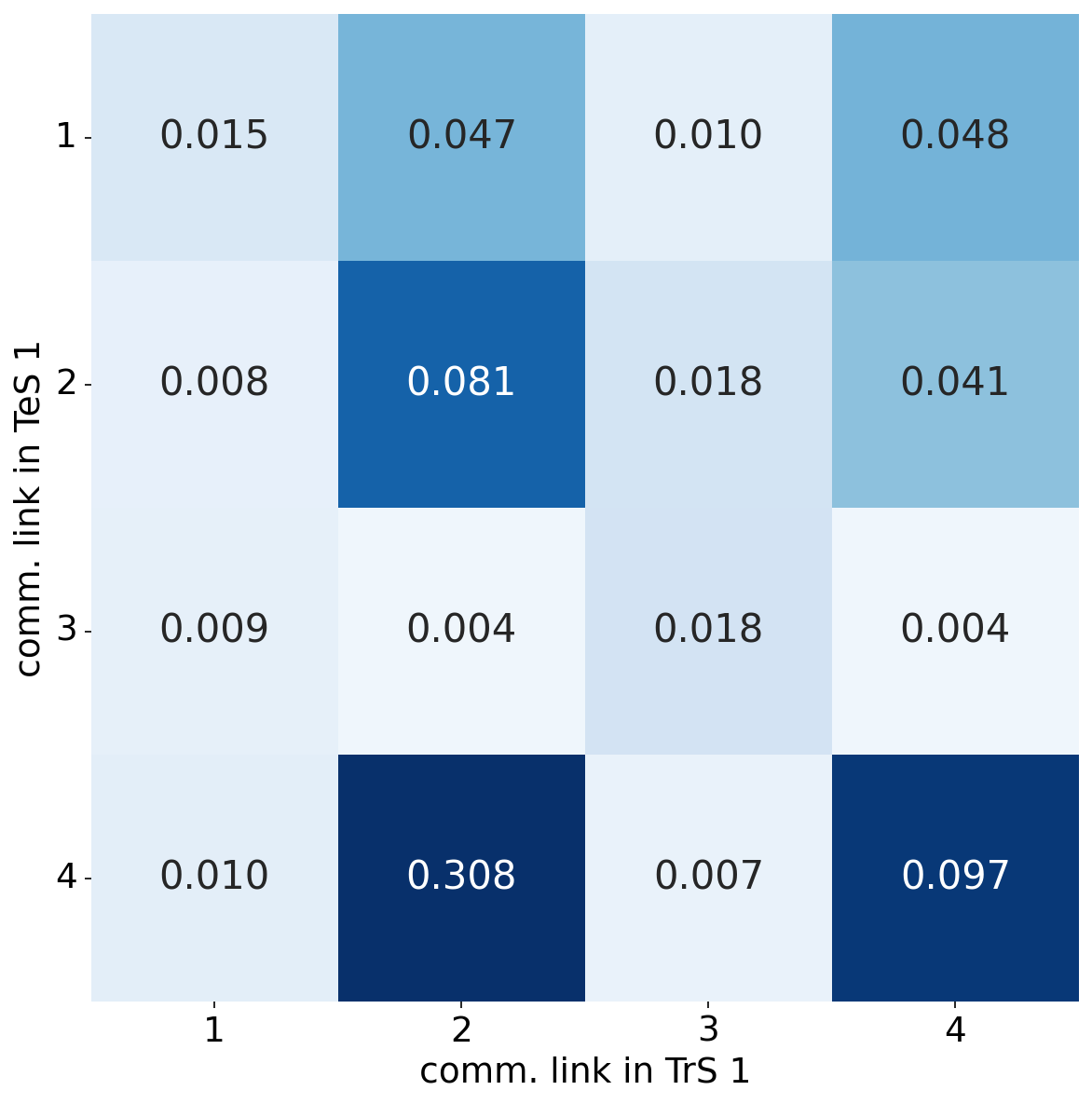}
  \caption{}
\end{subfigure}
\begin{subfigure}[t]{0.48\linewidth}
  \includegraphics[width=\linewidth]{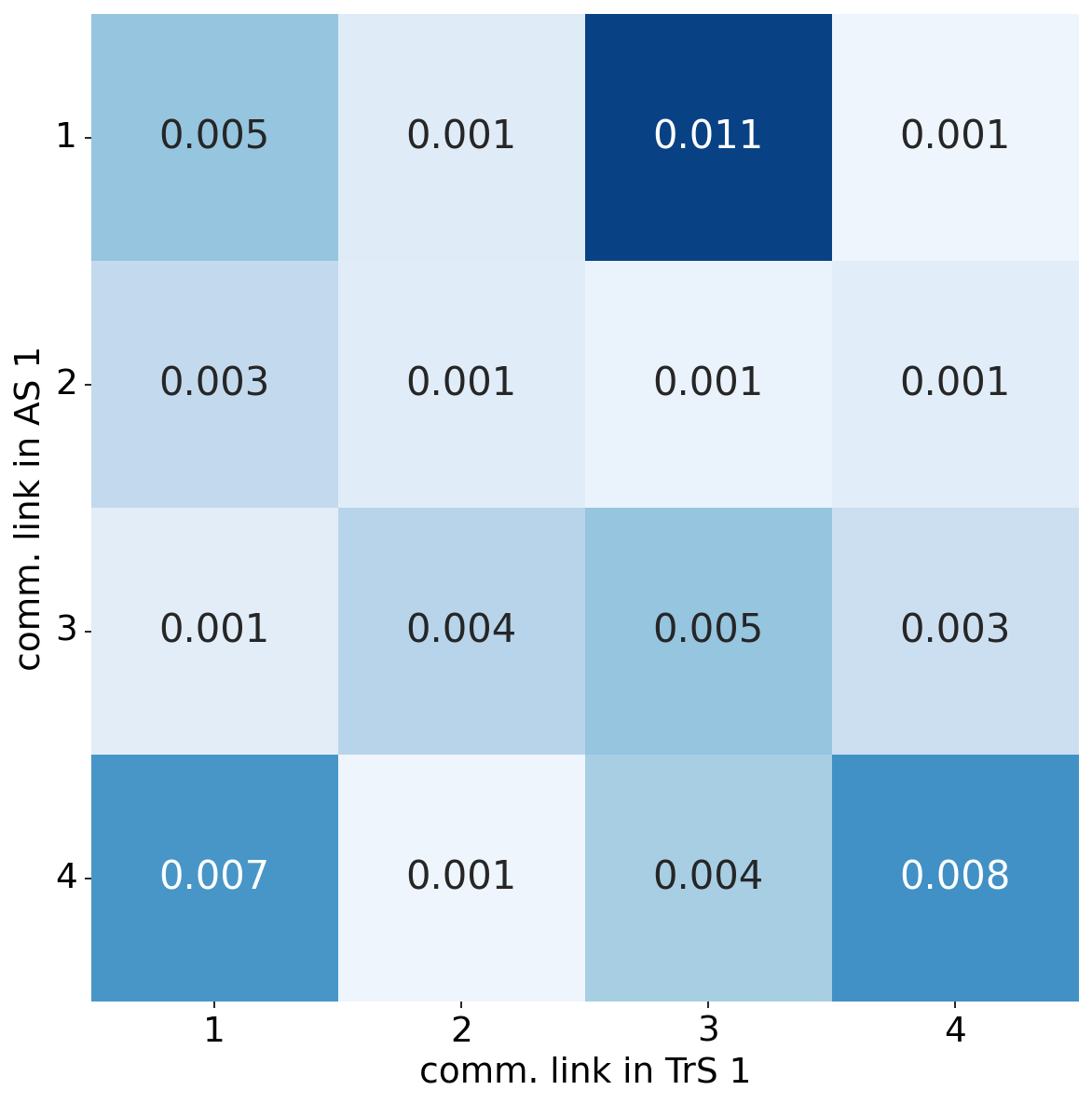}
  \caption{}
\end{subfigure}
\vspace{-5pt}
\caption{Correlation coefficient matrix for each communication pair: (a) between TrS 1 \& TeS 1; (b) between TrS 1 \& AS 1.}
\vspace{-10pt}
\label{fig:corr}
\end{figure}

\begin{figure}[t]
\centering
\begin{subfigure}[t]{0.48\linewidth}
  \includegraphics[width=\linewidth]{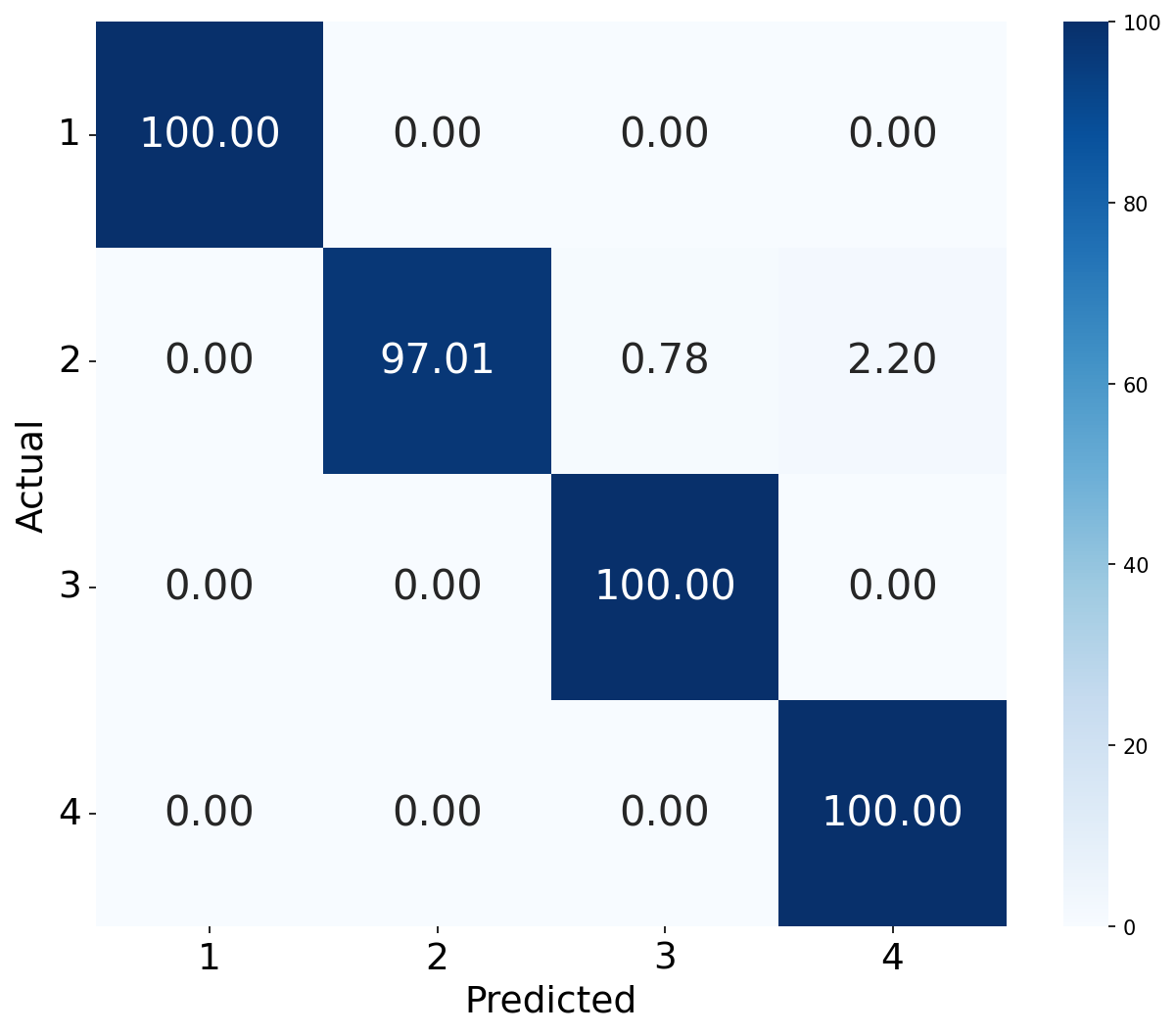}
  \caption{}
\end{subfigure}
\begin{subfigure}[t]{0.48\linewidth}
  \includegraphics[width=\linewidth]{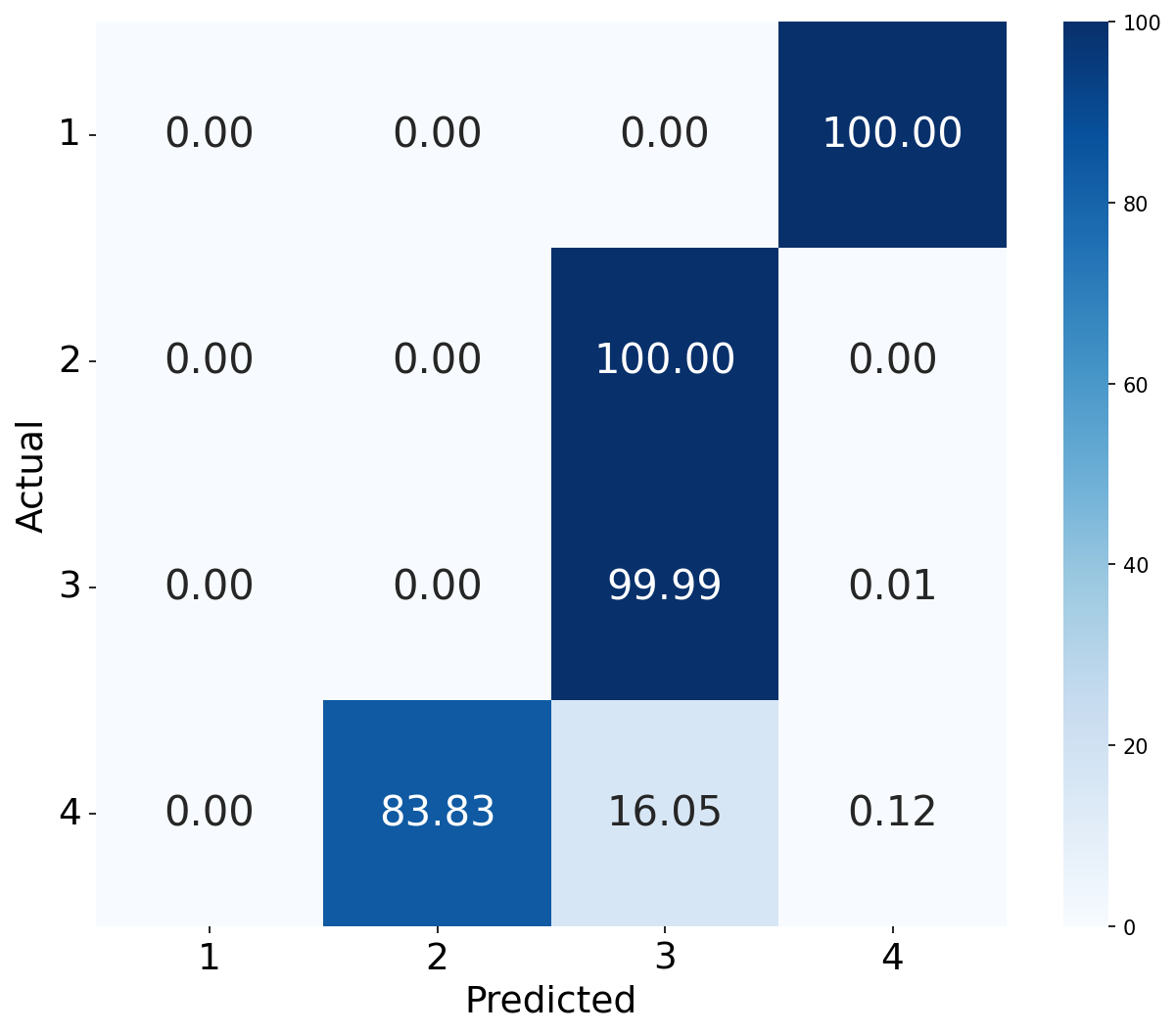}
  \caption{}
\end{subfigure}
\caption{Confusion matrix showing the classification accuracy of the CNN model trained on TrS 2: (a) tested on TeS 2; (b) tested on AS 2.}
\vspace{-10pt}
\label{fig:ex_same}
\end{figure}

We then consider the case when the CNN is trained with some publicly known, fixed, or low-entropy signals such as preambles, pilots, and headers and train the CNN with TrS 2. Surprisingly, when all the transmitters repeatedly transmit the same frame, the CNN model demonstrates robust and high classification accuracy despite domain shifts, as shown in Fig. \ref{fig:ex_same}(a), even when the signals are collected at significantly different times and locations, and the CSI in the training and testing data is temporally and spatially distinct, similar to that in Fig. \ref{fig:corr}.

Does this mean we can simply train the CNN to learn RF fingerprints from raw IQ samples by having transmitters send a fixed signal? Unfortunately, the results in Fig. \ref{fig:ex_differ}(b) suggest that the CNN is not learning pure RF fingerprints. Instead, some environmental and location-dependent information associated with the receiver, patterns of the transmitted signals, and the RF fingerprints are jointly preserved in the neurons. Regardless of the transmitter’s location, the signal received by the AP, which is fixed in a limited space such as a lab, is likely to pass through a limited number of scatterers, resulting in some common signal paths, especially considering that channels in typical multipath environments are dominated by 3–5 path components \cite{ghassemzadeh2004measurement,czink2007cluster}. Hence, for a transmitter whose RF fingerprint the CNN has learned, the corresponding neurons are activated, enabling the CNN to correctly identify the transmitter with high confidence even if the channel conditions change dramatically, as shown in Fig. \ref{fig:ex_same}(a).

On the other hand, for a transmitter with an RF fingerprint unknown to the CNN, the received signal still tends to contain common environmental characteristics associated with the AP used during training. As a result, the neurons preserving the environmental and location-dependent information of the AP are activated, causing the AP to misclassify the signal as originating from an authorized transmitter with high confidence, as shown in Fig. \ref{fig:ex_same}(b).

To further verify this insight, we evaluate the system's performance against a naive impersonation attack. We consider two scenarios in which Eve transmits either random or repeated signals, mimicking those of legitimate devices using her own transmitter. The success rates of these attacks are shown in Table \ref{tab:train5_15_testEve}. We observe that when Eve transmits repeated signals identical to those of TrS 2, which the CNN model is trained on, the model continues to exhibit the consistent misclassification behavior. In other words, despite Eve's RF fingerprints being distinct from those of the authorized transmitters, the model frequently misclassifies her signals as originating from a certain legitimate device. This misclassification is likely due to the model having learned not only RF fingerprints but also environmental and signal-pattern characteristics associated with the AP and transmitted frames, which dominate the model's performance. In contrast, when Eve performs the naive impersonation attack using random signals, the highest misclassification rate significantly drops from 98.06\% to 48.03\%,  further confirming that the transmitted signal pattern plays a key role in the model’s robustness and security.

\begin{table}[h]
    \centering
    \vspace{-5pt}
    \caption{Success rates of naive impersonation attacks}
    \begin{tabular}{|c|c|c|c|c|}
    \hline
          \multirow{2}{*}{Actual: Eve}& \multicolumn{4}{c|}{Predicted} \\
         \cline{2-5}
           & 1 & 2 & 3 & 4\\
           \hline
           repeated signals &  1.86\% & 0 & 0.09\% & 98.06\% \\ 
           \hline
           random signals & 13.8\% & 0.49\% & 48.03\% & 37.68\% \\
           \hline
    \end{tabular}    
    \label{tab:train5_15_testEve}
    \vspace{-5pt}
\end{table}

{\bf Remark 2.} {\em When the transmitted frame is deterministic, the CNN model trained with raw IQ samples can partially learn RF fingerprints associated with transceivers. However, it also jointly learns environmental and location-dependent information associated with the receiver, as well as patterns in the transmitted signals. These entangled features within the trained CNN model provide significant attack opportunities for both untargeted and naive impersonation attacks.}

\subsubsection{Thresholding CNN with softmax confidence is insufficient}
A straightforward approach to enhance the security of the CNN-based device identification system is to apply a threshold on the model’s softmax confidence scores, allowing it to reject signals from Eve as unknown. However, our study shows that this method does not improve the system’s resilience to replay or naive impersonation attacks, nor does it increase robustness against domain shifts.

\begin{table}[h]
    \centering
    \vspace{-5pt}
    \caption{Success and rejection rates of naive impersonation attacks against the CNN with softmax confidence}
    \begin{tabular}{|c|c|c|c|c|c|}
    \hline
          \multirow{2}{*}{Actual: Eve} & \multirow{2}{*}{Rejection} & \multicolumn{4}{c|}{Predicted}  \\
         \cline{3-6}
           & & 1 & 2 & 3 & 4 \\
           \hline
           repeated signals & 21.19\% & 0.44\% & 0 & 0.04\% & 99.52\% \\
           \hline
           random signals & 5.13\% & 11.77\% & 0.27\% & 52.4\% & 35.56\% \\
           \hline              
    \end{tabular} 
    \vspace{-5pt}
    \label{tab:train5_15_testEve_softmax}
\end{table}

Table \ref{tab:train5_15_testEve_softmax} presents the success and rejection rates of naive impersonation attacks when a confidence threshold of 0.95 is applied to the CNN’s output class probabilities via softmax. When the highest softmax score falls below this threshold, the CNN rejects the input as unknown. Comparing the results from Tables \ref{tab:train5_15_testEve} and \ref{tab:train5_15_testEve_softmax}, we observe that although the CNN correctly rejects some attacks, the overall rejection rate remains low (21.19\% when the transmitted signal pattern does not match, and 5.13\% when Eve transmits repeated signals). The CNN’s performance on TeS 2 and AS 2 with softmax confidence is similar to that shown in Fig. \ref{fig:ex_same}, with a low rejection rate in both cases. We omit the detailed results here due to page limitations. Consequently, the model’s performance remains nearly unchanged, as the jointly learned environmental and location-dependent information, along with transmitted signal patterns, still provide significant attack opportunities for Eve. 

{\bf Remark 3.} {\em The vulnerability caused by the lack of carefully designed signal preprocessing techniques for RF fingerprinting, combined with the limited capacity of neural networks to learn from complex and nonlinear wireless signals, means that applying post-processing security techniques such as a softmax confidence threshold to the CNN alone is insufficient to secure the system.}

\section{Discussion}
Our experimental findings necessitate the design of RF fingerprint estimation methods that can effectively disentangle device-specific hardware signatures from spatiotemporal channel and environmental artifacts. Special structures can be leveraged for estimating certain RF fingerprints, such as IQ imbalance, in a manner similar to how the CFO is estimated in WiFi standards using repeated symbols in synchronization words. Exploring hybrid models that integrate signal processing-based RF fingerprinting techniques with powerful DL architectures, such as embedding hardware impairment models or using generative models to simulate and enhance RF fingerprints, is also promising for providing robust and secure RF fingerprinting techniques. In our future work, we will design new signal processing techniques for RF fingerprinting and carefully integrate them with DL models to ensure system robustness and security.

\section{Conclusion}
We revisited CNN-based RF fingerprinting for device identification from a security perspective through an adversarial-driven experimental study. We demonstrated that although CNN models exhibit superior classification performance, they consistently show misclassification behaviors due to limitations in generalizing across temporal and spatial domain shifts in wireless signals. Attackers can exploit these behaviors to compromise the system through simple replay and naive impersonation attacks. Moreover, training CNNs with raw IQ samples causes the models to entangle RF fingerprints with environmental and signal-pattern features, creating additional attack vectors that cannot be mitigated by applying post-processing security techniques to the CNN alone.

\bibliographystyle{IEEEtran}
\bibliography{rf_attack}

\end{document}